\documentclass[aps,prb,twocolumn,10pt,superscriptaddress,showpacs]{revtex4-1}
\usepackage{epsfig}
\usepackage[dvips]{color}
\usepackage{bm}

\begin{document}

\title{Cluster expansion of multicomponent ionic systems with controlled accuracy: Importance of long-range interactions in heterovalent ionic systems}
\author{Atsuto \surname{Seko}}
\email{seko@cms.mtl.kyoto-u.ac.jp}
\affiliation{Department of Materials Science and Engineering, Kyoto University, Kyoto 606-8501, Japan}
\author{Isao \surname{Tanaka}}
\affiliation{Department of Materials Science and Engineering, Kyoto University, Kyoto 606-8501, Japan}
\affiliation{Nanostructures Research Laboratory, Japan Fine Ceramics Center, Nagoya 456-8587, Japan}

\date{\today}

\begin{abstract}
We have been examining factors determining the accuracy of cluster expansion (CE), which is used in combination with many density functional theory (DFT) calculations.
With the exception of multicomponent metallic or isovalent ionic systems, the contributions of long-range effective cluster interactions (ECIs) to configurational energetics are not negligible, which is ascribed to long-range electrostatic interactions.
The truncation of ECIs in such systems leads to systematic errors.
A typical problem with such errors can be seen in Monte Carlo (MC) simulations since simulation supercells composed of a larger number of atoms than those of the input DFT structures are used.
The prediction errors for long-period structures beyond the cell size of the input DFT structures in addition to those for short-period structures within the cell size of the input DFT structures need to be carefully examined to control the accuracy of CE.
In the present study, we quantitatively discuss the contribution of the truncation of long-range ECIs to the accuracy of CE.
Two types of system, namely, a point-charge spinel lattice and a real $ {\rm MgAl_2O_4}$ spinel crystal, are examined.
The prediction error of the long-period structures can be improved both by increasing the number of pairs and by also considering the effective screened electrostatic energy. 
\end{abstract}

\maketitle

\section{Introduction}

A combination of density functional theory (DFT) calculation and the cluster expansion (CE) method\cite{CE1,CE2,CE3} is widely used to evaluate thermodynamic properties, ground state structures and phase diagrams in multicomponent systems. 
The CE method gives an effective expression for the configurational energy.
In the CE method, the configurational energetics of a multicomponent system is characterized only using the effective cluster interactions (ECIs). 
To obtain accurate thermodynamic properties and phase diagrams at finite temperatures, it is essential to estimate the ECIs while controlling the prediction errors for the whole range of structures because the accuracy of the configurational density of states determines the thermodynamic properties.\cite{sampling:seko}

Within the formalism of the CE, the configurational energy $E$ of a binary system is expressed by using the pseudospin configurational variable $\sigma_i$ for the respective lattice site $i$ as
\begin{eqnarray}
E &=& V_0 + \sum\limits_{i} {V_i \sigma_i} + \sum\limits_{i,j} {V_{ij} \sigma_i \sigma_j} + \sum\limits_{i,j,k} {V_{ijk} \sigma_i \sigma_j \sigma_k} + \cdots \nonumber \\
&=& \sum\limits_{\alpha} V_\alpha \cdot \varphi_\alpha , 
\label{hamiltonian}
\end{eqnarray}
where $V_\alpha$ and $ \varphi_\alpha $ are called the ECI and correlation function of cluster $\alpha$, respectively.
Since only the atomic configuration determines the correlation functions, the ECIs characterize the configurational energetics of the binary system.
The unknown ECIs are generally estimated from the total energies of periodic ordered structures computed by DFT calculations using the least-squares technique.\cite{CE2} 

\begin{figure}[tbp]
\begin{center}
\includegraphics[width=\linewidth]{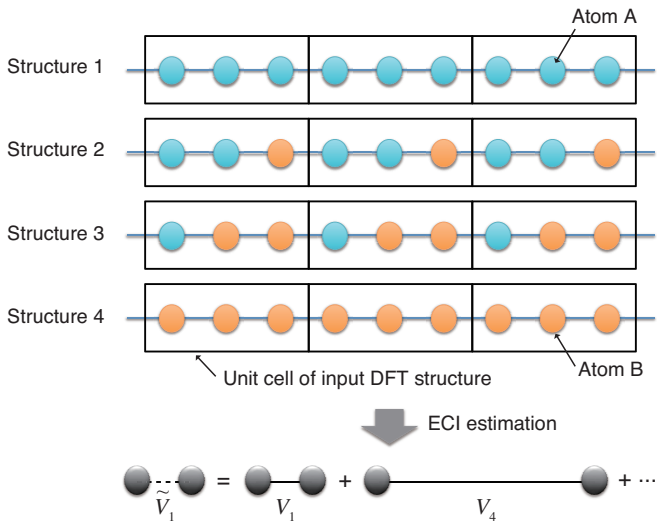} 
\caption{
Schematic illustration of DFT structures and ECI estimation in one-dimensional binary A-B system.
All input DFT structures have a periodicity of three atoms. 
For these input DFT structures, the correlation function of the first nearest neighbor (NN) pair is exactly the same as those of the fourth and seventh NN pairs.
Since the correlation functions of these pairs are mutually dependent, only one of the ECIs of the mutually dependent pairs can be estimated from the DFT total energies of the three-atom input structures.
If we choose the first NN pair for the ECI estimation, the contributions of the fourth and seventh NN pairs are incorporated into the estimated ECI of the first NN pair.
The estimated ECI of the first NN pair, $\tilde V_1$, is expressed using the true ECIs of the first, fourth and seventh NN pairs, $V_1$, $V_4$ and $V_7$, as $\tilde V_1 = V_1 + V_4 + V_7 + \cdots$.
When the ECIs of the fourth and seventh NN pairs are not negligible, they should be explicitly estimated by computing the DFT energies of longer-period structures.
}
\label{spinel_longrange:schematic_longrangeCE}
\end{center}
\end{figure}

In general, estimated ECIs deviate from the true values. 
One of the reasons for this is the truncation of long-range ECIs beyond the cell size of the input DFT structures.
It is natural to use a set of periodic ordered structures with a small number of atoms as the input set of DFT structures.
When the input DFT structures are expressed using a small number of atoms, only short-range ECIs of independent clusters within the cell size are estimated and the long-range ECIs beyond the cell size are truncated.
This situation can be easily seen in a one-dimensional binary system with only pairwise interactions as shown in Fig. \ref{spinel_longrange:schematic_longrangeCE}.
Meanwhile, the long-range interactions are taken into account in the DFT calculations through the periodicity of the wave functions and nuclear charges.
They should be associated with long-range ECIs beyond the cell size in the CE when they are not negligible.

The only use of short-range ECIs is expected to be appropriate for the CE of most multicomponent metallic systems and multicomponent ionic systems with configurations of only isovalent ions.
In multicomponent ionic systems with configurations of heterovalent ions, the contributions of long-range ECIs to the configurational energetics are not negligible, which is ascribed to the long-range electrostatic interactions.
CE using only short-range ECIs in such systems leads to systematic errors. 
A typical problem with such errors can be seen in Monte Carlo (MC) simulations to evaluate of thermodynamic properties since simulation supercells with a large number of atoms are used, in which the long-range ECIs play an important role.

In this study, we give a detailed discussion of the truncation of long-range ECIs by performing CEs in multicomponent ionic systems with configurations of heterovalent cations.
The strong dependence of the prediction error on the truncation of long-range ECIs is demonstrated.
In addition, we propose a procedure for optimizing the CE for the whole range of structures in systems with configurations of heterovalent ions.
We first apply the CE to the energetics for cationic configurations in a model system described only with point charges on the spinel structure. 
This model system will hereafter be called the "point-charge spinel lattice".
The CE is then applied to the cation energetics in a real $\rm MgAl_2O_4$ spinel.
Finally, the cationic order-disorder behavior in $\rm MgAl_2O_4$ is evaluated using the CE with a controlled accuracy for the whole range of structures.

The remainder of this paper is organized as follows. 
In Sec. \ref{spinel_longrange:method}, a general procedure for optimizing the CE is demonstrated. 
In Sec. \ref{spinel_longrange:coulomb_spinel_lattice}, an application of the CE to the point-charge spinel lattice is given.
The prediction errors for the configurational electrostatic energy and the order-disorder behavior are then examined.
In Sec. \ref{spinel_longrange:MgAl2O4}, the CE is applied to the cationic configurational properties of $\rm MgAl_2O_4$.
An optimal prediction of the order-disorder behavior in $\rm MgAl_2O_4$ is then given in Sec. \ref{spinel_longrange:MgAl2O4_manybody}.
Finally, we summarize this paper in Sec. \ref{spinel_longrange:summary}.

\section{Estimation of ECIs}
\label{spinel_longrange:method}
Let the configurational energy be composed of a finite number of terms, $m$.
When the energies of $N_{\rm DFT}$ $(> m)$ structures are computed by the DFT calculation, the ECIs for $m$ clusters are generally determined by the least-squares technique.\cite{CE2}
The least-squares technique minimizes
\begin{equation}
\sum\limits_{n = 1}^{N_{\rm DFT}} \left( \sum\limits_{\alpha} V_\alpha \cdot \varphi_\alpha^{(n)} - E_n \right)^2 , 
\end{equation}
where $E_n$ and $ \varphi_\alpha^{(n)}$ denote the DFT energy and correlation functions of cluster $\alpha$ for structure $n$, respectively.
In this framework, the accuracy of the CE is controlled by the number of clusters, $m$, the combination of clusters, the number of DFT structures, $N_{\rm DFT}$, and the combination of DFT structures.
These factors should be optimized as the configurational properties for the whole range of structures can be predicted within the accuracy of the DFT calculations.

\begin{figure}[tbp]
\begin{center}
\includegraphics[width=0.7\linewidth]{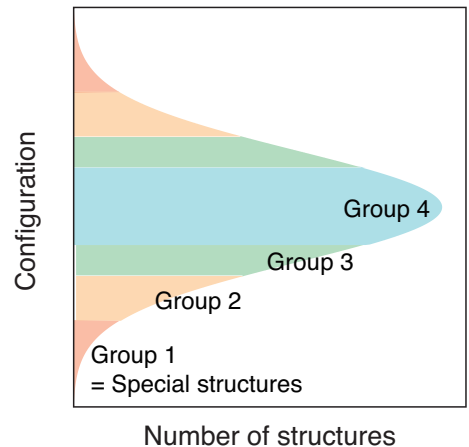} 
\caption{
Schematic illustration of distribution function of structure population in space of correlation functions.
}
\label{spinel_longrange:schematic_classification}
\end{center}
\end{figure}

To measure and control the accuracy of the CE for a wide range of structures, a procedure based on cluster analysis of the structure population (CASP) was proposed.\cite{casp:seko}
CASP classifies structures with similar correlation functions into the same group. 
Figure \ref{spinel_longrange:schematic_classification} shows a schematic illustration of the distribution function of the structure population in the configurational space.
In the schematic illustration, the structure population is classified into four groups according to the correlation functions.
Group 4 has the largest number of structures, which correspond to structures near a random structure.
On the other hand, Group 1 has the smallest number of structures, which are far from a random structure.
They will hereafter be called "special structures".
These special structures generally have correlation functions with large values.
Since the ground state structures are included in the special structures in many cases,\cite{Hart_Nature2} they are expected to be more important than the other structures.

The cross validation (CV) score,\cite{CV1,CV2} which is one of the scores used to evaluate the predictive power of the CE, has been widely used to control the accuracy of the CE.
The leave-one-out CV score is defined as
\begin{equation}
{\rm (CV)}^2 = \frac{1}{N_{\rm DFT}}\sum\limits_{n = 1}^{N_{\rm DFT}} {\left( {\hat E_{(n)}  - E_n} \right)^2 } ,
\label{cluster:cv}
\end{equation}
where $\hat E_{(n)}$ is the energy of structure $n$ predicted by the CE without using the DFT energy of structure $n$.
The number of clusters and their combination are optimized on the basis of the CV score.
The combination of clusters is obtained by minimizing the CV score for a fixed number of clusters using an efficient minimization algorithm such as the genetic algorithm.\cite{geneAlgo1,geneAlgo2}
However, the CV score is simply an average quantity for the input DFT structures and the errors of special structures are only a minor part of the CV score.
To calculate configuration-dependent properties at finite temperatures, for example, the configurational free energy as a function of temperature, the prediction errors over the whole range of configurations including the ground state and all excited states should be carefully examined.

Following CASP, we can estimate the accuracy for a wide range of structures including special structures by introducing the individual CV score for each group.
The CV score in group $\xi$ after CASP, which is denoted by CV-CASP$^{(\xi)}$, is expressed as
\begin{equation}
{\rm (CV \mathchar`- CASP }^ {(\xi)})^2 = \frac{1}{N_{\rm DFT}^{(\xi)}}\sum\limits_{n = 1}^{N_{\rm DFT}^{(\xi)}} {\left( {\hat E_{(n)}  - E_n} \right)^2 } ,
\label{spinel_longrange:cvca}
\end{equation}
where $N_{\rm DFT}^{(\xi)}$ denotes the number of DFT structures belonging to group $\xi$.
When the prediction error for special structures is larger than the average error, the accuracy for special structures can be improved by the simultaneous optimization of CE for all groups.
The simplest way to control the accuracy for all groups is to sample of DFT structures from each group evenly.
When DFT structures are selected from each group evenly, the square of the CV score corresponds to the average of the squares of the CV-CASPs.

\begin{figure}[tbp]
\begin{center}
\includegraphics[width=\linewidth]{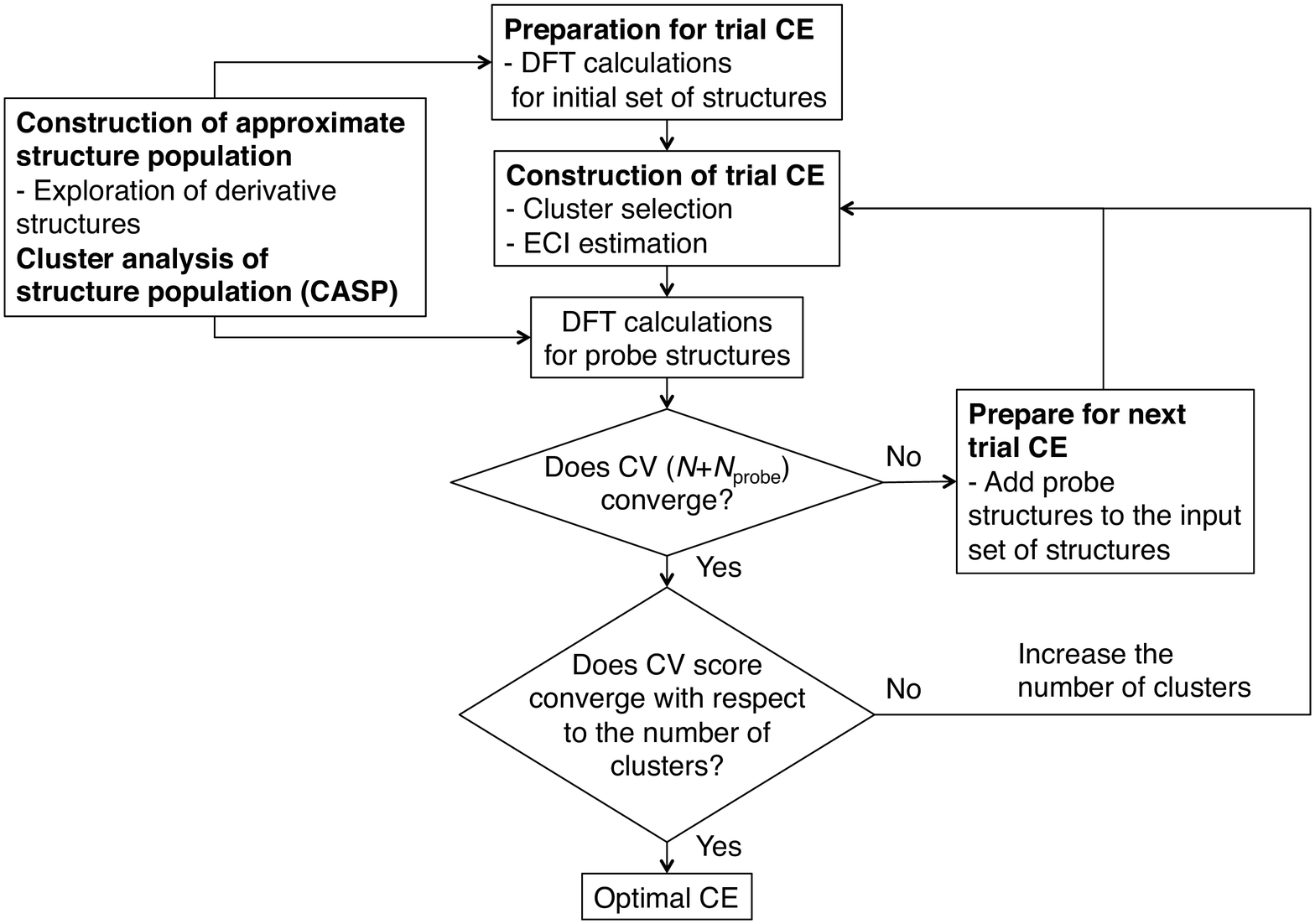} 
\caption{
Flowchart of the procedure for constructing the optimal CE based on CASP.
}
\label{spinel_longrange:flowchart}
\end{center}
\end{figure}

However, the combination of the uniform sampling of DFT structures and minimization of the CV score does not necessarily guarantee accuracy for the whole range of structures.
It is most important to validate a trial CE, which is constructed using a uniformly sampled input structure set by minimizing the CV score, by using another uniformly sampled structure set.
The additional structures are here called ”probe structures”.
When a trial CE is not optimal, the CV score and CV-CASPs are much smaller than the prediction error of the CE.
To avoid this underestimation, structures and clusters are alternately selected.\cite{sampling:seko,casp:seko}
A flowchart of the iterative procedure is drawn in Fig. \ref{spinel_longrange:flowchart}.


%

The procedure for optimizing the CE is composed of seven steps as follows.
(1) An approximate structure population is first prepared.
A set of derivative structures\cite{Hart_derivativestructure,Hart_derivativestructure2} is one of the candidates for the approximate structure population.
(2) The correlation functions for all structures in the structure population are calculated. 
CASP is then carried out according to the correlation functions.
(3) The initial DFT structures are selected from each group evenly and their DFT energies are computed.
(4) A trial CE with a fixed number of $m$ clusters is constructed by exploring the set of clusters that minimizes the CV score.
(5) The trial CE is validated using probe structures since the minimized CV score is often an underestimate of the error of the trial CE when the number of DFT structures is rather small with respect to the number of clusters.
$N_{\rm probe}$ probe structures are chosen from each group evenly and randomly.
The DFT energies of the probe structures are then calculated.
(6) The CV score for $N_{\rm DFT} + N_{\rm{probe}}$ structures using clusters selected from $N_{\rm DFT}$ structures in step (4), i.e., CV$(N_{\rm DFT} + N_{\rm{probe}})$, is evaluated to verify the convergence of the CE.
If CV$(N_{\rm DFT} + N_{\rm{probe}})$ is larger than the CV score evaluated in step (4), CV$(N_{\rm DFT})$, the trial CE is regarded as a failure.
In such a case, the probe structures are added to the input set. 
A new trial CE is then made in step (4) using $N_{\rm DFT} + N_{\rm{probe}}$ structures.
If CV$(N_{\rm DFT} + N_{\rm{probe}})$ is almost the same as CV$(N_{\rm DFT})$, the trial CE is regarded as optimal for the fixed number of $m$ clusters.
At the same time, an optimal set of DFT structures for the $m$ clusters is obtained.
Then, the construction of the CE proceeds to step (7) so as to optimize the number of clusters.
(7) By increasing the number of clusters and starting a new iterative step from step (4), the convergence of the CV score with respect to the number of clusters is examined.
This is repeated until the CV score converges with respect to the number of clusters.
Finally, the ECIs with the optimal number of clusters are estimated.
Using this iterative procedure, the number of clusters and their combination, the number of DFT structures and their combination are optimized.

\section{Cluster expansion on point-charge spinel lattice}
\label{spinel_longrange:coulomb_spinel_lattice}

\subsection{Structure of cation sublattice in spinel compounds}

A spinel compound with cations A and B and anion C has a general formula of $\rm{AB_2C_4}$, where the anions C form a nearly face-centered cubic (fcc) close-packed sublattice.
The spinel structure has two types of cation site, which are tetrahedral fourfold-coordinated and octahedral sixfold-coordinated sites.
The number of octahedral sites is double than that of the tetrahedral sites.
When all the tetrahedral sites are occupied by cation A, the spinel is called "normal", and when all the tetrahedral sites are occupied by cation B, the spinel is called "inverse".
The normal and inverse spinels are generally expressed using the formulae $\rm{A[B_2]C_4}$ and $ \rm{B[AB]C_4}$, respectively, where the square brackets indicate the octahedral sites.
The cation distribution on the tetrahedral and octahedral sites is characterized by the degree of inversion $x$, defined as the fraction of cations B on the tetrahedral sites or that of cations A on the octahedral sites.
The degree of inversion ranges from 0 (normal spinel) to 1 (inverse spinel).

\subsection{Point-charge spinel lattice}
We first apply the CE method to the configurational behaviors of cations on a point-charge spinel lattice.
Here we consider the configurations of cations A and B in the AB$_2$C$_4$ system with the spinel structure in which the energetics is described only by the electrostatic interactions among ions.
Here, ions A, B and C are substituted by point charges of $q_{\rm A} = +2$, $q_{\rm B} = +3$ and $q_{\rm C} = -2$, respectively.
The unit-cell shape of the point-charge spinel lattice is kept cubic. 
The lattice constant and internal parameter of the spinel are fixed to 8 \AA\ and 0.3855, respectively.
The electrostatic energy $E_{\rm es}$ for a configuration is expressed by
\begin{equation}
E_{\rm es} = \frac{1}{2} \sum_{i,j} \frac{q_i q_j}{r_{ij}},
\end{equation}
where $q_i$ and $r_{ij}$ denote the charge of ion $i$ and the distance between ions $i$ and $j$, respectively.
The electrostatic energy is evaluated by the Ewald method.
We use the $clupan$ code\cite{clupan,ame_feature:Seko,casp:seko} for the electrostatic energy calculations and subsequent CE constructions and MC simulations.

\subsection{CE construction}
First, all independent structures derived from the primitive cell of the spinel are exhaustively explored using the algorithm in Ref. \onlinecite{Hart_derivativestructure2}.
Table \ref{spinel_longrange:derivative} shows the numbers of independent structures.
Since the primitive cell is composed of six cations, the numbers of cations in the cells are multiples of six.
It is natural to use the set of independent structures as the approximate structure population.
Input structures are chosen from the independent structures.
Two different sets of input structures are prepared.
One is composed of all the structures with up to 12 cations (28 atoms), which will hereafter be called "12-cation input structures".
The number of structures is $3 + 75 = 78$.
The other is composed of all the structures with up to 18 cations (42 atoms), which will hereafter be called "18-cation input structures".
The total number of structures is $3 + 75 + 2288 = 2366$.
The computed electrostatic energies for the 78 and 2366 structures are used as the input energies.

\begin{table}[tbp]
\caption{Numbers of independent structures along with the numbers of primitive cells and cations.}
\label{spinel_longrange:derivative}
\begin{ruledtabular}
\begin{tabular}{ccc}
Number of & Number of & Number of \\
cells & cations & independent structures \\
\hline
1 & 6 & 3 \\
2 & 12 & 75 \\
3 & 18 & 2288 \\
4 & 24 & 149644 \\
5 & 30 & 4080208
\end{tabular}
\end{ruledtabular}
\end{table}

The ECIs are then estimated using two kinds of input set of energies.
Since the electrostatic interaction is a pairwise interaction, the cation-configuration energetics of the point-charge spinel lattice can be expressed using the ECIs of clusters only up to pair clusters.
The number of independent clusters whose ECIs can be estimated is dependent on the number of atoms included in the input structures.
When 12- and 18-cation input structures are used, 14 and 42 ECIs are independent, respectively.
Here, CEs with up to the maximum number of clusters are made using the input electrostatic energies.
The empty and point clusters are considered in all the CEs.

\begin{figure}[tbp]
\begin{center}
\includegraphics[width=\linewidth]{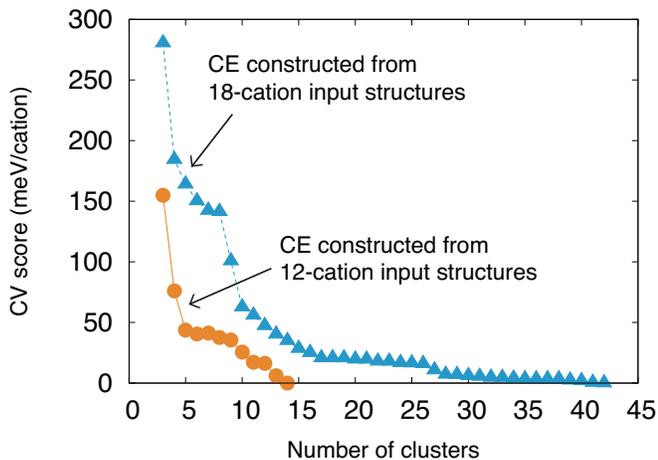} 
\caption{
Dependence of the CV score on the number of clusters on the point-charge spinel lattice.
}
\label{spinel_longrange:point_charge_cv}
\end{center}
\end{figure}

Figure \ref{spinel_longrange:point_charge_cv} shows the dependence of the CV score on the number of clusters.
Pair clusters are included in the CE according to their pair distance.
The CV scores decrease monotonically as the number of clusters increases.
The CV score becomes zero at $m=14$ for the 12-cation input structures, where all independent clusters are considered, whereas it is $m=42$ for the 18-cation input structures.
The CV score of the CE (12-cation) is lower than that of the CE (18-cation) for the same number of clusters.
This is natural since the CV score of the CE (12-cation) is simply the prediction error only for structures with up to 12-cation structures.
The CE (12-cation) does not have predictive power for longer-period structures.
Therefore, the absolute value of the CV score is not an appropriate measure for the prediction error.

%

\subsection{Predictive power}
Next, we discuss the predictive powers for structures that are not considered in the input set.
The prediction errors of CEs for 12-, 18-, 24- and 30-cation structures are estimated.
The prediction errors for the 12- and 18-cation structures are estimated from all the 12- and 18-cation independent structures, respectively. 
The prediction errors for the 24- and 30-cation structures are estimated from a set of 100 randomly selected independent structures.
The root-mean-square (RMS) differences between the CE and electrostatic energies for the structure sets are regarded as the prediction errors.

\begin{figure}[tbp]
\begin{center}
\includegraphics[width=\linewidth]{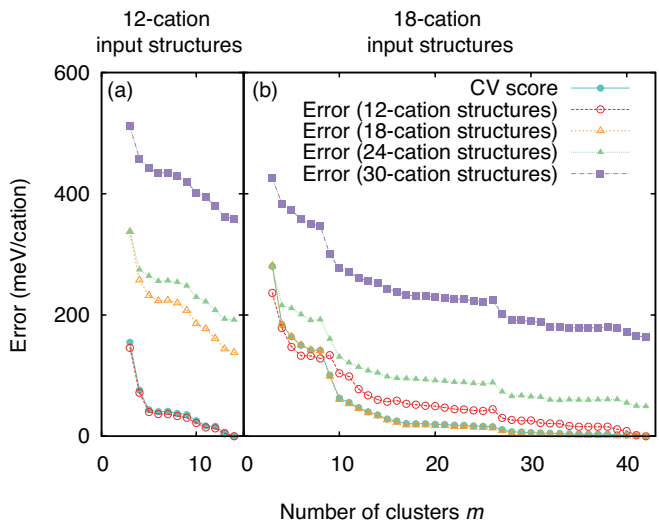} 
\caption{
Prediction errors of CEs constructed from (a) 12-cation input structures and (b) 18-cation input structures for 12-, 18-, 24- and 30-cation structures.
The CV scores are also shown for comparison.
}
\label{spinel_longrange:point_charge_prediction}
\end{center}
\end{figure}

Figure \ref{spinel_longrange:point_charge_prediction} shows the prediction errors of CEs constructed from (a) 12-cation input structures and (b) 18-cation input structures for 12-, 18-, 24- and 30-cation structures.
The prediction error decreases as the number of clusters increases.
Compared with the prediction errors for structure sets with different numbers of cations in each CE, the prediction error increases with the number of cations included in the structures used for the prediction.
In other words, the contribution of the truncation of long-range ECIs to the prediction error becomes larger with increasing number of cations included in the structures used for prediction.
Exceptionally, the prediction error for the 12-cation structures are larger than that for the 18-cation structures in the CEs (18-cation).
The CEs (18-cation) are excessively optimized for the 18-cation structures at the expense of the accuracy for the 12-cation structures.

As can be seen in Fig. \ref{spinel_longrange:point_charge_prediction}, the prediction errors for the 12- and 18-cation structures are very close to the CV scores of CEs (12-cation) and CEs (18-cation), respectively.
In the CE of $m=14$ (12-cation) with the CV score of exactly zero, the prediction error for the 12-cation structures becomes zero.
Similarly, in the CE of $m=42$ (18-cation) with the CV score of exactly zero, the prediction errors for both the 12- and 18-cation structures are exactly zero.
The electrostatic energy of a structure with up to 18 cations can be expressed using 42 ECIs with no errors.
On the other hand, the prediction errors for the 24- and 30-cation structures are much larger than the CV scores.
Although the CV scores are exactly zero in the CEs of $m=14$ (12-cation) and $m=42$ (18-cation), the prediction errors for the 24- and 30-cation structures are completely different from the CV scores.
As also described above, the CV score is not suitable for estimating the prediction error for longer-period structures than the input structures.

However, the behaviors of the CV score and the prediction errors with respect to the number of clusters are reasonably similar.
This implies that the minimization procedure for the CV score is useful for minimizing the prediction error for long-period structures.
Therefore, when practically constructing an accurate CE for both short- and long-period structures, a trial CE is first made from the input structures by minimizing the CV score.
Then, the trial CE should be validated using longer-period probe structures than the input structures to guarantee the accuracy of the long-period structures.
If the prediction error for the longer-period probe structures is not acceptable, the prediction error should be improved by including the longer-period structures in the input structures and increasing the number of pair clusters.
This can be iteratively repeated until the prediction error for structures with a longer period than the input structures becomes acceptable.

\subsection{Order-disorder behavior}

We next discuss the contributions of the prediction error to the order-disorder behavior on the point-charge spinel lattice.
The temperature dependence of the cation distribution on the point-charge spinel lattice is predicted from two types of CE with $m = 5$ and $m=14$ constructed from 12-cation input structures and four types of CE with $m=10$, $m=16$, $m=28$ and $m=42$ constructed from 18-cation input structures.
The ground state structure is first explored by calculating the energies of all independent structures with up to 30 cations from the CEs.
As a result, the normal spinel configuration has the lowest energy among all six types of CE.
Finite-temperature thermodynamic properties are then evaluated using canonical MC simulations. 
The normal spinel is adopted as the initial structure in the MC simulations.
The MC simulations are performed up to 15000 K at temperature intervals of 1000 K.
Supercells for the MC simulations are constructed by the $20 \times 20 \times 20$ expansion of the primitive cell, which contain 48000 cations.
The MC simulations are performed over 2000 MC steps per cation to calculate the thermodynamic averages after equilibration over 5000 MC steps per cation.

\begin{figure}[tbp]
\begin{center}
\includegraphics[width=\linewidth]{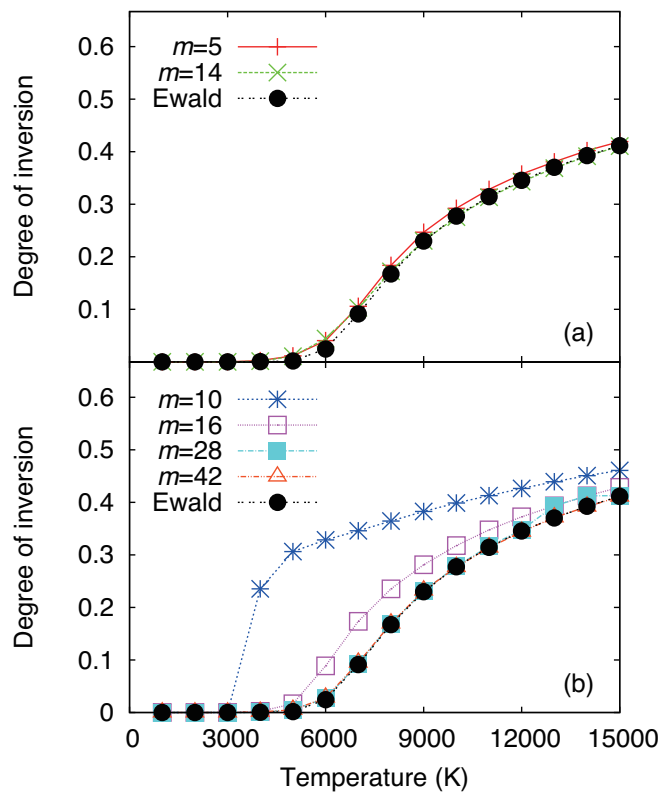} 
\caption{
Temperature dependences of the degree of inversion calculated using (a) two types of CE constructed from 12-cation input structures and (b) four types of CE constructed from 18-cation input structures on a point-charge spinel lattice.
The exact temperature dependence of the degree of inversion calculated from MC simulations in which configurational electrostatic energies are exactly computed by the Ewald method is also shown by closed circles for comparison.
}
\label{spinel_longrange:point_charge_inversion}
\end{center}
\end{figure}

Figure \ref{spinel_longrange:point_charge_inversion} shows the temperature dependences of the degree of inversion calculated using the (a) two types of CE (12-cation) and (b) four types of CE (18-cation).
The temperature dependence of the degree of inversion calculated from MC simulations in which configurational electrostatic energies are exactly computed by the Ewald method is also shown in Fig. \ref{spinel_longrange:point_charge_inversion}.
The degrees of inversion predicted using the CEs (12-cation) show coincidental agreement with the exact values despite the CEs of $m=5$ and $m=14$ both having large prediction errors for long-period structures as shown in Fig. \ref{spinel_longrange:point_charge_prediction} (a).
On the other hand, the degrees of inversion predicted using the CEs of $m=10$ (18-cation) and $m=16$ (18-cation) are inconsistent with the exact values.
This may be ascribed to the fact that the prediction errors for the 12-cation structures of these CEs (18-cation) are larger than those of the CEs (12-cation) as shown in Fig. \ref{spinel_longrange:point_charge_prediction} (b).
To improve the prediction of the order-disorder behavior, it is necessary to include a larger number of pair clusters. 
The prediction of the order-disorder behavior may also be improved by controlling the accuracy for a wide range of structures using a weighted optimization on the basis of CASP.
Here, the predicted degree of inversion becomes closer to the exact value by including a larger number of pair clusters, as can be seen in Fig. \ref{spinel_longrange:point_charge_inversion} (b).
To obtain accurate finite temperature thermodynamic properties, it is essential to estimate and control the prediction errors for both short-period and long-period structures.

The findings obtained from CE calculations on the point-charge spinel lattice are summarized as follows.
In systems with configurations of heterovalent ions, the CV score is the predicted error for short-period structures with periods having the same magnitude as the input structures.
The CV score cannot be used as the predicted error for longer-period structures.
However, the minimization of the CV score is still valid for optimizing the CE because the CV score and prediction error for long-period structures are mutually dependent.
To guarantee the accuracy for long-period structures, a trial CE constructed by the minimization of the CV score should be validated using long-period probe structures.
Since the increase in the accuracy for long-period structures sometimes leads to a decrease in the accuracy for short-period structures, the prediction errors for both short- and long-period structures should be examined.

\section{Cluster expansion of M\lowercase{g}A\lowercase{l$_2$}O\lowercase{$_4$}}
\label{spinel_longrange:MgAl2O4}
\subsection{DFT calculations}
We hereafter demonstrate the CE on the cation sublattice in a real MgAl$_2$O$_4$ spinel.
CASP is first carried out.
By selecting input structures evenly and randomly from each classified group, it is expected that estimate the accuracy for all structures can be estimated in an unbiased manner.
As an approximate structure population, the set of independent structures with up to 18 cations shown in Table \ref{spinel_longrange:derivative} is adopted.
The total number of structures in the structure population is 2366.
Here CASP is performed by model-based cluster analysis.\cite{FraleyRaftery2002,MclustSoftwareManual2006}
The likelihood of the correlation functions of all structures in the structure population is modeled by a Gaussian mixture.
CASP is performed according to the correlation functions of 117 independent clusters up to the ninth NN quadruplets.
The likelihood is maximized using the expectation-maximization (EM) algorithm for each of the 100 types of Gaussian mixture model.
We regard the model with the lowest Bayesian information criterion (BIC)\cite{bic} among the 100 models as the best one.
In the best model, the structure population is divided into five groups. 
The numbers of structures in the five groups are shown in Table \ref{spinel_longrange:casp_n_structure}.

\begin{table}[tbp]
\caption{Numbers of structures belonging to groups classified by CASP.}
\label{spinel_longrange:casp_n_structure}
\begin{ruledtabular}
\begin{tabular}{cc}
Group index & Number of structures \\
\hline
1 & 570 \\
2 & 434 \\
3 & 369 \\
4 & 407 \\
5 & 586 
\end{tabular}
\end{ruledtabular}
\end{table}

We prepare an input set of DFT structures to construct the CE.
The input set is composed of 250 structures containing the 150 structures with up to 18 cations selected evenly and randomly from each classified group and 100 structures with 24 cations selected randomly from the set of 24-cation independent structures.
The structures with 24 cations are included in the input set so as to consider a large number of pair clusters.
In addition to the input DFT structures, another three sets of DFT structures are prepared in order to estimate the prediction error for structures that are not included in the input set.
The three sets are (1) 100 structures with up to 18 cations selected evenly and randomly from the classified group, (2) 100 structures selected randomly from all 24-cation independent structures and (3) 100 structures selected randomly from all 30-cation independent structures.
The prediction error is estimated from the RMS difference between the CE energies and DFT energies.
DFT calculations are performed by the projector augmented-wave (PAW) method\cite{PAW1,PAW2} within the local density approximation (LDA)\cite{LDA1,LDA2} as implemented in the VASP code.\cite{VASP1,VASP2}
The plane-wave cutoff energy is set to 350 eV.
The total energies converge to less than 10$^{-2}$ meV.
The atomic positions and lattice constants are relaxed until the residual forces become less than $10^{-2}$ eV$/$\AA.

\subsection{CE with clusters up to pairs}
\begin{figure}[tbp]
\begin{center}
\includegraphics[width=\linewidth]{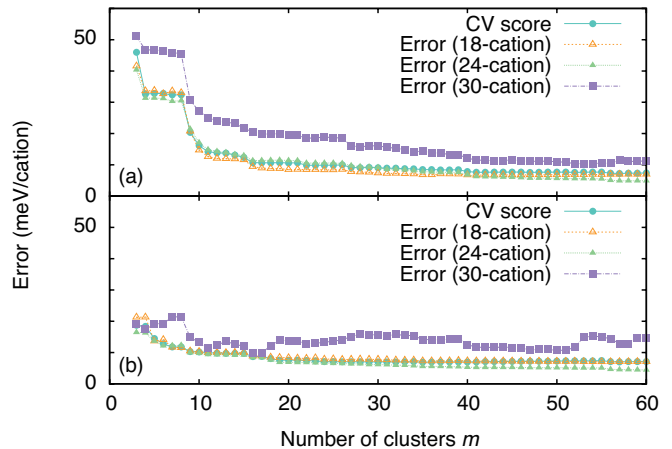} 
\caption{
(a) CV scores of CEs with clusters up to pairs constructed from 250 structures with up to 24 cations.
The prediction errors for 18-, 24- and 30-cation structures of the CEs are also shown.
(b) CV scores of CEs with clusters up to pairs combined with a screened point-charge model.
The prediction errors for 18-, 24- and 30-cation structures are also shown.
}
\label{spinel_longrange:spinel_prediction}
\end{center}
\end{figure}

Using the input DFT energies, the CE is carried out using only clusters up to pairs following the procedure described in Sec. \ref{spinel_longrange:coulomb_spinel_lattice}.
The ECIs are estimated by least-squares fitting from the energies and correlation functions of the input DFT structures.
Figure \ref{spinel_longrange:spinel_prediction} (a) shows the CV score and prediction errors for the 18-, 24- and 30-cation structures.
Pair clusters are included according to the distance of the pair.
The dependence of the CV score on the number of clusters is similar to that obtained from the CEs on the point-charge spinel lattice shown in Fig. \ref{spinel_longrange:point_charge_cv}.
This implies that the energetics of $\rm MgAl_2O_4$ is mostly Coulombic.
The CV score appears to converge at $m_{\rm pair} = 16$.
The prediction errors for the 18- and 24-cation structures are almost the same as the CV score because input set contains both 18- and 24-cation structures.
However, the prediction error for the 30-cation structures is much larger than the CV score even though the CV score converges.
To decrease the prediction error for the 30-cation structures, a larger number of clusters should be used regardless of the convergence of the CV score.

%

To improve the CE for long-period structures by procedures other than increasing the number of pair clusters, a description of the configurational long-range interaction may be needed.
We therefore adopt a combination of the CE and a screened point-charge model (SPCM).
This approach is used in Ref. \onlinecite{vandeWalle_CeO2-Sm}.
To distinguish the normal CE from the combined model, we hereafter call the combined model "CE-SPCM".
Within the CE-SPCM, the configurational energy $E$ of a binary system is expressed by adding a term describing the effective screened electrostatic energy, given by
\begin{equation}
E = \sum\limits_{\alpha} V_\alpha \cdot \varphi_\alpha + \frac{1}{2 \varepsilon} \sum_{i,j} \frac{q_i q_j}{r_{ij}},
\label{hamiltonian_ewald}
\end{equation}
where $\varepsilon$ is a screening parameter and the point charges of $q_i$ and $q_j$ are set to $q_{\rm Mg} = +2$, $q_{\rm Al} = + 3$ and $q_{\rm O} = -2$.
The first term is the general CE formulation and the second term is the effective screened electrostatic energy.
Short-range and long-range interactions are described by the CE and the effective screened electrostatic energy, respectively.
Since the correlation functions and electrostatic energy depend on only the atomic configuration, the ECIs $V_\alpha$ and screening parameter $\varepsilon$ are simultaneously optimized by minimizing
\begin{equation}
\sum\limits_{n = 1}^{N_{\rm DFT}} \left| \sum\limits_{\alpha} V_\alpha \cdot \varphi_\alpha^{(n)} + \frac{1}{2 \varepsilon} \sum_{i,j} \frac{q_i q_j}{r_{ij}} - E_n \right|^2 
\end{equation}
using the least-squares technique.

CE-SPCMs are carried out using the input set of DFT structures.
Figure \ref{spinel_longrange:spinel_prediction} (b) shows the CV score and prediction errors for the 18-, 24- and 30-cation structures of CE-SPCMs.
Pair clusters are included in the CE according to the pair distance.
When the number of pair clusters is small ($m_{\rm pair} < 15$), both the CV score and prediction errors are smaller than those for the normal CE because the contributions of truncated ECIs to the energetics are large.
When the number of pair clusters is $m_{\rm pair} \geq 15$, the CV score is not improved from the normal CE.
On the other hand, the prediction error for the 30-cation structures is improved from the normal CE by including the effective screened electrostatic energy.
This means that the inclusion of the effective screened electrostatic energy is useful for improving the predictions for longer-period structures than the input structures. 

However, it should be noted that the predictions are not necessarily improved by using the CE-SPCM.
The prediction error for 30-cation structures is scattered with respect to the number of pair clusters in the CE. 
In the CE-SPCM, the screening parameter is mainly determined from the contributions of clusters that are not included in a truncated form of the CE.
When only a small number of independent clusters are used for the CE, a screening parameter that is meaningful for long-period structures can be obtained.
When a large number of independent clusters are used for the CE, the screening parameter must be determined only from the contributions of a small number of clusters that are not included in the truncated form.
In such a case, the obtained effective screened electrostatic energy cannot be applied to long-period structures.
If the prediction error for long-period structures is significantly scattered with respect to the number of clusters and/or much larger than the CV score in the whole range of the number of clusters, it is necessary to include longer-period structures in the input set.

In the practical construction of an optimal CE, it is more essential to control both the CV score and the prediction error for longer-period structures than the input structures, as also described in the application to the point-charge spinel lattice.
As can be seen in Figs. \ref{spinel_longrange:point_charge_prediction} and \ref{spinel_longrange:spinel_prediction}, the behaviors of the prediction errors are similar regardless of the number of cations included in the structures used for prediction.
In other words, the prediction errors for structure sets with different numbers of cations are dependent on each other.
Therefore, it is expected that the prediction errors for structures with longer periods than 30 cations will be small in a CE with a small prediction error for 30-cation structures.
Here it may be sufficient to optimize the CE for only 30-cation structures as structures with longer periods than the input structures.
Since the CV score and prediction errors for 18- and 24-cation structures almost converge to small values at $m_{\rm pair} = 10$, we regard the CE with the lowest prediction error for 30-cation structures as the optimal CE among the CEs with $m_{\rm pair} \geq 10$.
As can be seen in Fig. \ref{spinel_longrange:spinel_prediction}, the CE-SPCM with $m_{\rm pair}=16$ is here regarded as the optimal pair CE.
In the optimal pair CE, the prediction errors for 18-, 24- and 30-cation structures are 8.8, 8.4 and 9.8 meV/cation, respectively.
In this optimal pair CE, the prediction error for structures with longer periods than the input structures is almost the same as those for structures in the input set.

\section{Order-disorder behavior in M\lowercase{g}A\lowercase{l$_2$}O\lowercase{$_4$}}
\label{spinel_longrange:MgAl2O4_manybody}

\begin{figure}[tbp]
\begin{center}
\includegraphics[width=\linewidth]{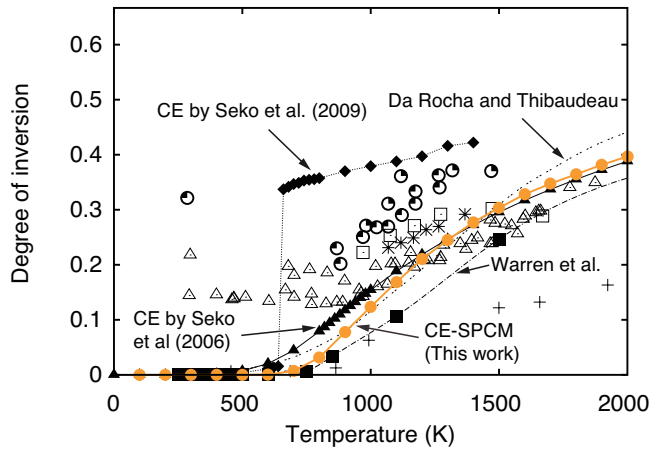} 
\caption{
Temperature dependence of the degree of inversion $x$ calculated from the optimal CE-SPCM of $m=42$, shown by closed orange circles.
The temperature dependences of the degree of inversion calculated using simple approximations with a few parameters for the internal energy\cite{MgAl2O4:Warren1,MgAl2O4:Rocha} and CEs with short-range ECIs\cite{MgAl2O4:Seko,sampling:seko} are also shown by lines and closed symbols.
Experimental values obtained by neutron diffraction measurements,\cite{MgAl2O4:disorder:neutron1,MgAl2O4:disorder:neutron2} x-ray diffraction measurements\cite{MgAl2O4:disorder:x-ray1,MgAl2O4:Tc2} and NMR\cite{MgAl2O4:disorder:NMR1,MgAl2O4:disorder:NMR2} are also shown by open symbols.
}
\label{spinel_longrange:spinel_inversion}
\end{center}
\end{figure}

$\rm{MgAl_2O_4}$ spinel is well known to have a normal cation configuration in the ground state. 
As the temperature increases, $\rm{MgAl_2O_4}$ undergoes disordering, resulting in the exchange of atoms on the tetrahedral and octahedral sites as can be observed using many experimental techniques.\cite{MgAl2O4:disorder:neutron1,MgAl2O4:disorder:neutron2,MgAl2O4:disorder:x-ray1,MgAl2O4:Tc2,MgAl2O4:disorder:NMR1,MgAl2O4:disorder:NMR2}
Computational approaches using a combination of DFT calculations and statistical mechanics techniques have also been carried out to examine the temperature dependence of the cation distribution.
Figure \ref{spinel_longrange:spinel_inversion} shows the computed temperature dependences of the degree of inversion along with experimental results.
Warren $et$ $al.$ applied canonical MC simulations to determine the temperature dependence of the degree of inversion using a few short-range interactions parameterized from DFT calculations for some ordered structures.\cite{MgAl2O4:Warren1,MgAl2O4:Warren2}
Da Rocha and Thibaudeau used a quadratic form of the internal energy with respect to the degree of inversion parameterized by fitting DFT energies to study the temperature dependence.\cite{MgAl2O4:Rocha}
The entropy for the disordered state with degree of inversion $x$ was analytically obtained by the point approximation.
In both studies, the computed temperature dependences coincide with the experimental ones at high temperatures in spite of the use of simple approximations.
We carried out two kinds of calculations based on a combination of DFT calculations, the CE method and MC simulations for investigating the temperature dependence of the degree of inversion.
In Ref. \onlinecite{MgAl2O4:Seko}, MC simulations were performed using five short-range pair ECIs and 17 many-body ECIs estimated from DFT calculations for 115 randomly selected ordered structures.
The predicted continuous behavior of the degree of inversion was close to those obtained by the two other calculations using simple approximations.\cite{MgAl2O4:Warren1,MgAl2O4:Rocha}
A more accurate CE with the smallest prediction error for structures that were not included in the input set of DFT structures among the previous calculations was made by optimizing the input set of DFT structures.\cite{sampling:seko}
This CE contained only short-range ECIs, similarly to the CE in Ref. \onlinecite{MgAl2O4:Seko}.
Although the prediction error of the CE in Ref. \onlinecite{sampling:seko} was the smallest, the temperature dependence of the degree of inversion showed discontinuous behavior in strong contrast to the other calculations.

We finally predict the order-disorder behavior in $\rm MgAl_2O_4$ using the optimal CE with many-body clusters based on the optimal pair CE.
Many-body clusters are selected from a set of candidate many-body clusters with pairs up to ninth NN.
The numbers of candidate clusters are shown in Table \ref{spinel_longrange:spinel_clusters}.
Simulated annealing is carried out to find the optimal set of many-body clusters with the minimum CV score instead using of the genetic algorithm.
The empty, point and 14 pair clusters used in the optimal pair CE are included in the cluster set.

\begin{table}[tbp]
\caption{
Numbers of candidate clusters on the cation lattice in the spinel.
Although two point clusters are symmetrically independent, one of the two point clusters is only independent when considering cation configurations in the fixed composition of ${\rm MgAl_2O_4}$.
}
\label{spinel_longrange:spinel_clusters}
\begin{ruledtabular}
\begin{tabular}{cc}
Number of lattice sites & Number of clusters \\
\hline
empty & 1 \\
1 & 2 \\
2 & 14 \\
3 & 32 \\
4 & 82 \\
5 & 121 \\
6 & 123 \\
\hline
total & 375
\end{tabular}
\end{ruledtabular}
\end{table}

\begin{figure}[tb]
\begin{center}
\includegraphics[width=\linewidth]{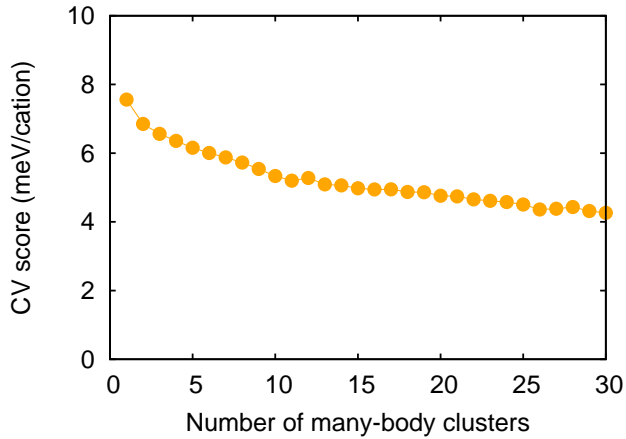} 
\caption{
Dependence of the CV score on the number of many-body clusters. 
The CV score is minimized by selecting a combination of many-body clusters for a fixed number of many-body clusters.
}
\label{spinel_longrange:spinel_cv_manybody}
\end{center}
\end{figure}

The optimal number of many-body clusters is explored by estimating the CV scores for different numbers of many-body clusters $m_{\rm MB}$ from $m_{\rm MB} = 1$ to $m_{\rm MB} = 30$.
Figure \ref{spinel_longrange:spinel_cv_manybody} shows the dependence of the CV score on $m_{\rm MB}$.
Since the CV score gradually decreases as $m_{\rm MB}$ increases, the CE for $m_{\rm MB} = 26$ is here adopted as the optimal CE.
The total number of clusters is $m=42$.
The CV score of the optimal CE is 4.5 meV/cation.
The prediction errors for 18-, 24- and 30-cation structures that are not included in the input DFT structures are 5.2, 4.5 and 10.3 meV/cation, respectively.
Although the prediction errors for the 18- and 24-cation structures are almost the same as the CV score, the prediction error for the 30-cation structures is larger than the CV score and close to that of the optimal pair CE of 9.8 meV/cation.
The prediction error for the 30-cation structures cannot be decreased to the same magnitude as those for the 18- and 24-cation structures because only many-body clusters with pairs up to ninth NN are considered here.

The temperature dependence of the degree of inversion is then calculated from canonical MC simulations using the optimal CE of $m=42$.
Figure \ref{spinel_longrange:spinel_inversion} also shows the calculated temperature dependence of the degree of inversion.
The degree of inversion calculated from the optimal CE-SPCM is close to that calculated from a simple approximation with only two interaction parameters carried out by Da Rocha and Thibaudeau.\cite{MgAl2O4:Rocha}
On the other hand, the degree of inversion calculated from the CE without long-range ECIs\cite{sampling:seko} is considerably different from that calculated from the optimal CE-SPCM.
The continuous change in the degree of inversion may be better expressed by the simple approximation than by using the CE without long-range ECIs.
Since only the accurate configurational density of states is required to predict the order-disorder behavior, the truncation of complex short-range and long-range interactions may be cancelled out. 
This result implies that it is essential to control the prediction errors for both short- and long-period structures.

\section{Summary}
\label{spinel_longrange:summary}
In the present study, we quantitatively discussed the relationship between ECI truncation and the predictive power of the CE in heterovalent ionic systems.
The CE was applied to two types of multicomponent ionic system: a point-charge spinel lattice and a real ${\rm MgAl_2O_4}$ spinel crystal.
We found that the CV score, which is widely adopted as a criterion for ECI truncation, is applicable only for evaluating the prediction errors of short-period structures within the cell size of input DFT structures.
The optimization of the CE based only on the CV score leads to systematic errors of long-period structures beyond the cell size of the input DFT structures.
Therefore, it is essential to control the prediction error of the long-period structures in addition to the CV score.
When the prediction error of the long-period structures is not acceptable, the CE should be optimized for both short- and long-period structures after including the long-period structures in the input set of DFT structures.
The prediction error for the long-period structures can be reduced by increasing the number of pairs and/or by also considering the effective screened electrostatic energy.
The simultaneous optimization of the CE for both short- and long-period structures enables us to accurately predict configurational thermodynamic properties and phase diagrams in multicomponent systems.

\begin{acknowledgments}
This study was supported by a Grant-in-Aid for Young Scientists (A) from Japan Society for the Promotion of Science (JSPS), Japan.
IT also acknowledgments supports in the form of both a Grant-in-Aid for Scientific Research (A) and a Grant-in-Aid for Scientific Research on Innovative Areas "Nano Informatics" (grant number 25106005) from JSPS.

\end{acknowledgments}

\bibliography{spinel}

\end{document}